\documentclass[doublecol]{epl2} 
\usepackage{amsmath,amssymb}

\title{Three dimensional contractile droplet under confinement}

\author{A. Tiribocchi\inst{1,2} \and M.  Lauricella\inst{1} \and A. Montessori\inst{3}\and S. Succi\inst{1,4,5}}
\shortauthor{F. Author \etal}

\institute{                    
  \inst{1} Istituto per le Applicazioni del Calcolo, Consiglio Nazionale delle Ricerche, via dei Taurini 19, Roma, 00185, Italy\\
  \inst{2} INFN "Tor Vergata" Via della Ricerca Scientifica 1, 00133 Roma, Italy\\
   \inst{3} Department of Civil, Computer Science and Aeronautical Technologies Engineering, Roma Tre University, via Vito
Volterra 62, Rome, 00146, Italy\\
  \inst{4} Center for Life Nano- \& Neuro-Science, Fondazione Istituto Italiano di Tecnologia, viale Regina Elena 295, 00161 Rome, Italy\\
  \inst{5} Department of Physics, Harvard University, 17 Oxford St, Cambridge, MA 02138, United States
}

\abstract{We numerically study the dynamics of a three-dimensional contractile fluid droplet in the bulk and under confinement. We show that varying activity leads to a variety of shapes and motile regimes whose motion is driven by an interplay between spontaneous flows and elasticity. In the bulk the droplet self-propels unidirectionally, acquiring either an almost spherical shape at intermediate activity or a peanut-like geometry for larger values. Under confinement, the droplet exhibits a previously unreported oscillating dynamics characterized by periodic hits against opposite walls of a microchannel while moving forward. These results could be of interest for the study of artificial microswimmers and their biological analogs, such as living cells.}

\begin{document}

\maketitle

\section{Introduction}

Active droplets are a bio-inspired class of soft material where self-propulsion stems from an energy supply located within the droplets or in their surroundings \cite{maass2,maass_epje,michelin,Weber_2019}. A particular sub-class is represented by active gel droplets where the driving force is generated by a hierarchically-assembled active liquid crystal of biological origin, such as suspensions of microtubules and kinesin or actomyosin solutions \cite{dogic,sagues2,sagues3}. 
The  former are an example of extensile materials, which pull the fluid inwards along the equator and emit it axially, while the latter are contractile where the fluid dynamics is reversed. 
These systems are of particular interest for studying the mechanics of cell locomotion \cite{tjhung1,tjhung3,yeom_prxlife} and for the design of artificial microswimmers, potentially useful in material science \cite{pablo,dogic3}.

While a large body of theoretical studies have been dedicated to modeling 
the physics of active gel droplets in two \cite{giomi,fialho,hawkins,sci_rep_bon,demagistris,aranson1,aranson2} and three dimensional setups \cite{tjhung1,tjhung2,tjhung3,ruske,yeom_prxlife,carenza_pnas,negro_natcomm}, much less is known about the physics under confinement \cite{tiribocchi_nat,tiribocchi_pof2023,tiribocchi_sm2023}, especially in three dimensions \cite{winkler_comm}. In this work, we move a step further and investigate the hydrodynamics of a contractile fluid droplet confined within a three dimensional microchannel, a geometry that could potentially reproduce the cell migration in a realistic (or physiological) environment.
Following previous works \cite{tiribocchi_nat,tiribocchi_pof2023}, we use a continuum approach incorporating hydrodynamics, where the advection-relaxation equations govern the evolution of a phase field, accounting for active material concentration, and a vector field describing its orientation, while the Navier-Stokes equations control the global velocity. Simulations in bulk, while confirming previous shapes and dynamic regimes \cite{tjhung1}, unveil a novel motile state where a peanut-like droplet hosting an integer topological defect (of charge $1$ or $-1$) self-propels. Under mild confinement, at sufficiently high activity a highly stretched droplet is found to exhibit a periodic motion characterized by repeated bumps against opposite walls. This dynamics is suppressed under a high confinement regime.

\section{Methods}
The hydrodynamic model used in this work follows those presented, for example, in \cite{tjhung2,tiribocchi_nat,tiribocchi_sm2023,tiribocchi_pof2023}. We consider a fluid mixture consisting of an active gel droplet immersed in a passive Newtonian fluid of constant density $\rho$. The droplet phase is described by a scalar field $\phi({\bf r},t)$, positive in the droplet and zero outside, while the ordering properties of the active material contained in the droplet are captured by a polar liquid crystal vector ${\bf p}({\bf r},t)$, representing a coarse-grained average of the orientation of the internal constituents (e.g an actin filament). Finally, the global fluid velocity (of droplet and solvent) is tracked by a vector field ${\bf v}({\bf r},t)$.

The dynamics of the phase field $\phi$ is governed by an advection-relaxation equation
\begin{equation}\label{eq_phi}
\partial_t\phi+\nabla\cdot({\bf v}\phi)
=M\nabla^2\mu,
\end{equation}
where $M$ is the mobility and $\mu=\frac{\delta{\cal F}}{\delta\phi}$ is the chemical potential, being ${\cal F}$ the free energy (see later in the text).

The equation of the polarization field ${\bf p}$ reads
\begin{equation}\label{eq_p}
 \partial_t{\bf p}+ {\bf v}\cdot\nabla{\bf p}
 = \underline{{\bf s}}\cdot{\bf p}-\frac{1}{\Gamma}{\bf h},
\end{equation}
where $\underline{{\bf s}}=\chi\underline{{\bf D}}-\underline{\mathbf{\Omega}}$, with $\underline{{\bf D}}=(\underline{{\bf W}}+\underline{{\bf W}}^T)/2$ and $\underline{{\mathbf \Omega}}=(\underline{{\bf W}}-\underline{{\bf W}}^T)/2$ being the symmetric and antisymmetric parts of the velocity gradient tensor $W_{\alpha\beta}=\partial_{\beta}v_{\alpha}$ (Greek indices denote Cartesian components). Also, the parameter $\chi$ is a shape factor, positive for rod-like particles and negative for disk-like ones, $\Gamma$ is the rotational viscosity and ${\bf h}=\frac{\delta {\cal F}}{\delta {\bf p}}$ is the molecular field governing the relaxation of the liquid crystal.

Fluid density $\rho$ and velocity ${\bf v}$ obey the continuity and Navier Stokes equations  which, in the limit of incompressible fluid, are
\begin{equation}\label{cont_eq}
\nabla\cdot{\bf v}=0,
\end{equation}
\begin{equation}\label{nav_stok}
\rho\left(\frac{\partial}{\partial t}+{\bf v}\cdot\nabla\right){\bf v}=-\nabla P + \nabla\cdot(\underline{\sigma}^{\textrm{active}}+\underline{\sigma}^{\textrm{passive}}),
\end{equation}
where $P$ is the isotropic fluid pressure. On the right hand side of Eq.(\ref{nav_stok})
 $\sigma_{\alpha\beta}^{active}=-\zeta\phi\left(p_{\alpha}p_{\beta}-\frac{1}{d}|{\bf p}|^2\delta_{\alpha\beta}\right)$, where $\zeta$ is the activity (positive for extensile materials and negative for contractile ones) and $d$ is the dimension of the system. Moreover, $\sigma_{\alpha\beta}^{passive}$ is the sum of three contributions, the dissipative $\sigma_{\alpha\beta}^{\textrm{viscous}}=\eta(\partial_{\alpha}v_{\beta}+\partial_{\beta}v_{\alpha})$, where $\eta$ is the viscosity, the elastic $\sigma_{\alpha\beta}^{\textrm{elastic}}=\frac{1}{2}(p_{\alpha}h_{\beta}-p_{\beta}h_{\alpha})-\frac{\chi}{2}(p_{\alpha}h_{\beta}+p_{\beta}h_{\alpha})-\kappa\partial_{\alpha}p_{\gamma}\partial_{\beta}p_{\gamma}$, where $\kappa$ is the elastic constant of the liquid crystal and the interfacial 
$\sigma_{\alpha\beta}^{\textrm{interface}}=\left(f-\phi\frac{\delta{\cal F}}{\delta\phi}\right)\delta_{\alpha\beta}-\frac{\partial f}{\partial(\partial_{\beta}\phi)}\partial_{\alpha}\phi$, where $f$ is the free energy density. 

The free energy ${\cal F}$ used to compute the thermodynamic forces (chemical potential, molecular field and stress tensor) appearing in Eqs.(\ref{eq_phi})-(\ref{eq_p})-(\ref{nav_stok}) can be written as an expansion up to the fourth order of $\phi$ and ${\bf p}$ capturing the bulk properties of the active gel plus gradient terms describing interfacial features of the fluid mixture and local distortions of the liquid crystal \cite{marchetti,tiribocchi3}. It is given by
\begin{eqnarray}\label{free}
{\cal F}&=&\int_VdV\biggl(\frac{a}{4\phi_{cr}^4}\phi^2(\phi-\phi_0)^2+\frac{k}{2}(\nabla\phi)^2\nonumber.\\&&-\frac{\alpha}{2}\frac{(\phi-\phi_{cr})}{\phi_{cr}}|{\bf p}|^2+\frac{\alpha}{4}|{\bf p}|^4+\frac{\kappa}{2}(\nabla{\bf p})^2\biggr),
\end{eqnarray}
where $a$ and $k$ are positive constants controlling surface tension $\sigma=\sqrt{\frac{8ak}{9}}$ and interface thickness $\xi=\sqrt{\frac{2k}{a}}$. The first term of Eq.(\ref{free}) allows for the coexistence of two minima, $\phi=\phi_0$ inside the droplet and $\phi=0$ outside, while the second one penalizes the formation of fluid interfaces. The other terms represent the liquid crystal contribution, where the first two, multiplied by the constant $\alpha$, describe the bulk properties (with $\phi_{cr}=\phi_0/2$ critical concentration at which the isotropic-polar transition occurs) and the third one accounts for the elastic deformations in the single elastic constant approximation \cite{degennes}.

\subsection{Numerical implementation and mapping to real units} 
We numerically solve the hydrodynamic equations using a hybrid lattice Boltzmann method \cite{succi,tiribocchi_prep,tiribocchi3}, where Eq.(\ref{cont_eq}) and Eq.(\ref{nav_stok})
are integrated using a standard LB approach while Eq.(\ref{eq_phi}) and Eq.(\ref{eq_p}) through a finite difference Euler scheme, where a stencil representation of the finite difference operators is employed to ensure isotropy and numerical stability \cite{thampi}.

Simulations are run in three dimensional boxes, where two parallel flat walls are placed at distance $L_z$ while periodic boundary conditions are set along the $x$ and $y$ directions. At the walls, we impose no-slip conditions for the fluid velocity (i.e. $v_{z=0,Lz}=0$) and no-wetting for the phase field (i.e. $\phi_{z=0,Lz}=0$), while no anchoring is set for the polarization field. Also, we impose no anchoring of {\bf p} at the fluid interface.
We initialize the simulation by placing, in the center of the box, a droplet of radius $R=15$ where $\phi({\bf r},0)\simeq 2$ and ${\bf p}({\bf r},0)={\bf p}_y({\bf r},0)$ (with $|{\bf p}|=1$) for $r\le R$, while $\phi$ and ${\bf p}$ are set to zero for $r>R$.
Also, we keep the radius constant and vary the distance between the walls, in order to change the confinement ratio $\lambda=L_z/2R$, and the activity $\zeta$.

The simulation parameters have been chosen as follows: $a=4\times 10^{-2}$, $k=6\times 10^{-2}$, $M=10^{-1}$, $\alpha=10^{-1}$, $\kappa=4\times 10^{-2}$, $\chi=1.1$, $\Gamma=1$, $\eta\simeq 1.67$, $\Delta x=1$ (lattice step) and $\Delta t=1$ (time step). These values yield 
$\sigma\simeq 4.5\times 10^{-2}$ and $\xi\simeq 2$.
Finally, the active parameter $|\zeta|$ ranges approximately from $10^{-3}$ to $10^{-2}$, values that guarantee a spontaneous motion with minimal chance of droplet breakup. 

A typical mapping between these parameters and real units can be found in Ref.\cite{tiribocchi_pof2023,tiribocchi_sm2023}. In brief, assuming that time, length and force scales are $T=10$ms, $L=1$$\mu$m and $F=100$nN and considering a simulation box of size $L_x=100$, $L_y=150$ and $L_z=50$, our system would roughly map a microfluidic channel of height $\simeq 50$$\mu$m,  hosting a droplet of diameter $D\simeq 30$$\mu$m with a shear viscosity 
$\simeq 1.5$kPa s and an elastic constant $\simeq 4$nN. Finally, the surface tension would approximately correspond to $4.5$mN/m and the activity ranges roughly between $1-10$ kPa. Note that the viscosity of the fluids inside and outside the droplet are set to equal values, an approximation that could be released in principle by letting $\phi$ depend on $\eta$.

\section{Results}

\subsection{Contractile droplet in unconfined geometry}

\begin{figure*}[htbp]
\includegraphics[width=1.0\textwidth]{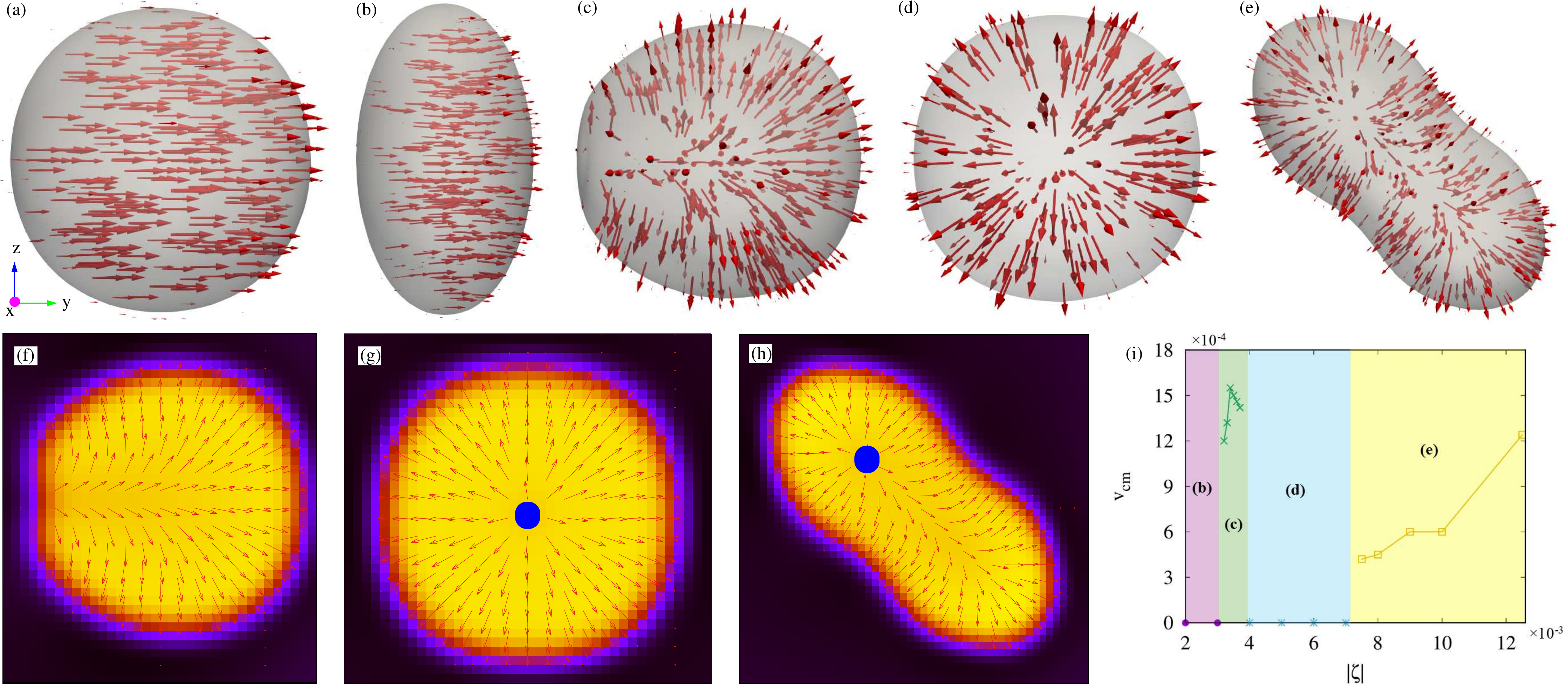}
\caption{(a) Equilibrium configuration of a passive fluid droplet ($\zeta=0$). Red arrows represent the polar field while the gray surface is a contour where $\phi\simeq 1$. These apply to all figures. (b-e) Steady state configurations of droplet and polar field for $|\zeta|=3\times 10^{-3}$ (b), $|\zeta|=3.7\times 10^{-3}$ (c), $|\zeta|=6\times 10^{-3}$ (d) and $|\zeta|=8\times 10^{-3}$ (e).  For low values of $\zeta$ the droplet attains an ellipsoidal non-motile shape where ${\bf p}$ remains essentially uniform (b) while, for intermediate values, the droplet turns almost spherical and motile. Here ${\bf p}$ exhibits a splay deformation (c). Increasing $\zeta$ leads to a non-motile spherical shape where ${\bf p}$ displays a hedgehog configuration (d) while, for higher values, a peanut-like structure emerges, where ${\bf p}$ acquires an asymmetric splay (e).
(f-h) Two dimensional section (in the y-z plane at $x=L_x/2$) of the polar field  for $|\zeta|=3.7\times 10^{-3}$ (a), $|\zeta|=6\times 10^{-3}$ (b) and  $|\zeta|=8\times 10^{-3}$ (c). Blue dots indicate topological defects of charge $+1$.
(i) Phase diagram of steady state center of mass speed $vs$ activity. Color stripes are associated with the different regimes.}
\label{fig1}
\end{figure*}

We begin by considering a three dimensional fluid droplet of diameter $D=30$ immersed in a passive Newtonian fluid and containing a polar liquid crystal initially oriented along the $y$ direction (see Fig.\ref{fig1}a). The size of the simulation box is $L_x=100$, $L_y=150$ and $L_z=105$ (large enough to minimize the interference of periodic images of the droplet) and  periodic conditions are set at the boundaries. Once the droplet is relaxed towards equilibrium, we turn on $|\zeta|$ and observe its dynamics over time. We confirm the existence of three different steady states already discussed in Ref.\cite{tjhung2,tjhung3}  and report evidence of  a new steady state, at high values of $|\zeta|$, previously undetected. 

\begin{figure*}[htbp]
\includegraphics[width=1.0\textwidth]{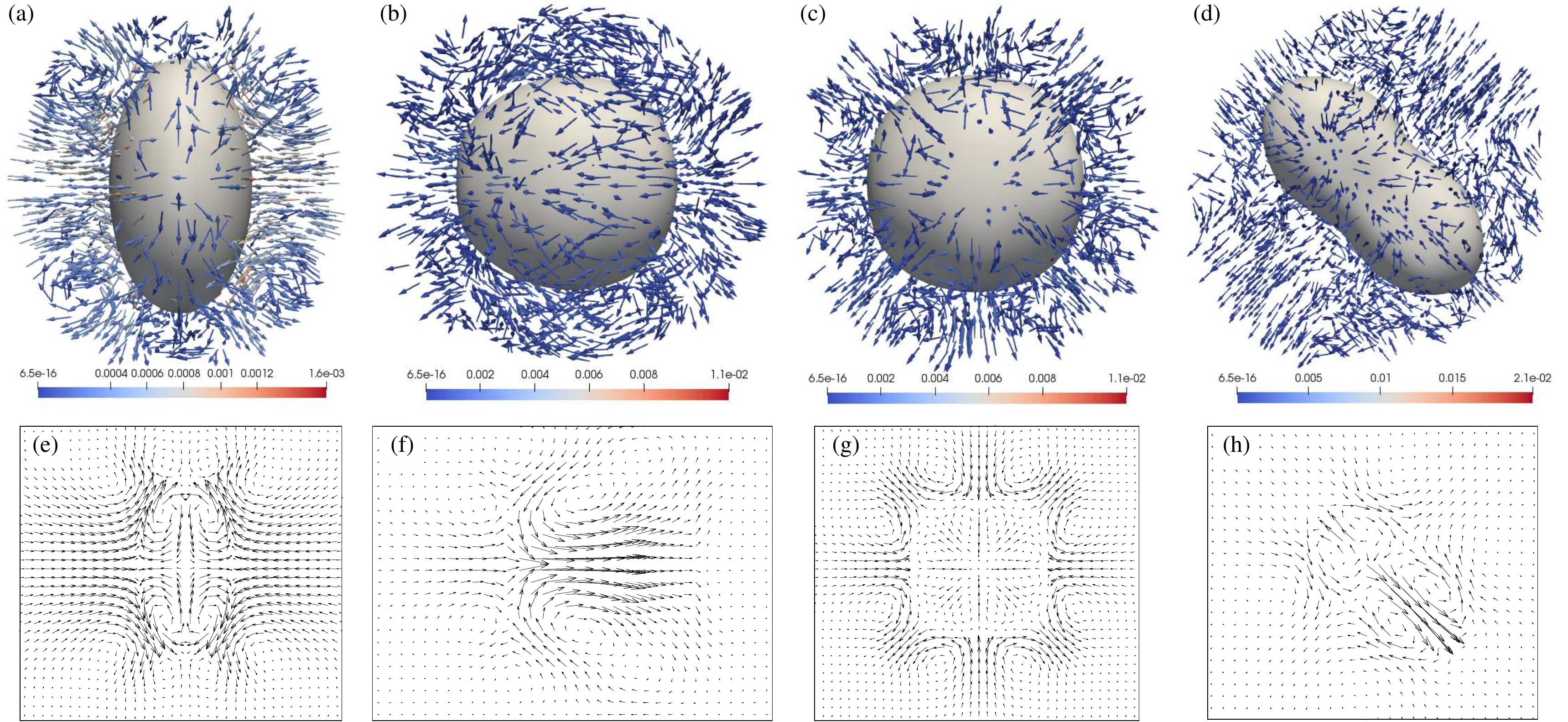}
\caption{Three dimensional structure of the velocity field in the surrounding of the droplet for (a) $|\zeta|=3\times 10^{-3}$, (b) $|\zeta|=3.7\times 10^{-3}$, (c) $|\zeta|=6\times 10^{-3}$ and (d) $|\zeta|=8\times 10^{-3}$. The color bar represents the magnitude of the velocity. Figures (e)-(f)-(g)-(h) represent the corresponding two dimensional section, taken at $x=x_{cm}$ in the $y-z$ plane.
While at low values of $|\zeta|$ (a-e) the droplet is non-motile because of a symmetric four-roll mill collecting the fluid equatorially and expelling it longitudinally, at intermediate values of $|\zeta|$ (b-f) the velocity exhibits a double-vortex pattern propelling the droplet horizontally. Increasing $|\zeta|$ leads to either a non-motile state where the velocity displays a symmetric eightfold structure or, for higher values, to a motile state where a jet of fluid, located within approximately half of the droplet, drives the motion.} 
\label{fig2}
\end{figure*}

For $10^{-3}\lesssim|\zeta|\lesssim 3\times 10^{-3}$,
the droplet acquires a steady elliptical shape (Fig.\ref{fig1}b) caused by a balance between contractile stress, which sets a four-roll flow in the droplet vicinity (Fig.\ref{fig2}a,e), and interfacial tension plus liquid crystal elasticity, opposing shape deformations. For values lower than $10^{-3}$ such changes are negligible.
A suitable dimensionless quantity gauging this balance is the active droplet number ${\cal A}=\frac{|\zeta| R}{\sigma +\kappa/R}$ which is approximately $0.95$  with $|\zeta| =3\times 10^{-3}$, $\sigma\simeq 4.5\times 10^{-2}$ and $\kappa/R\simeq 2.5\times 10^{-3}$. On a general basis, self propulsion  occurs if ${\cal A}>1$, whereas the droplet remains non-motile for ${\cal A}<1$.

Increasing $|\zeta|$ leads to a rectilinear motion, whose direction is set by $(\nabla\cdot{\bf p}){\bf p}$. If $3\times 10^{-3}<|\zeta|\lesssim 4\times 10^{-3}$ (here ${\cal A}$ ranges approximately between $1.1$ and $1.3$), the polarization becomes unstable to splay deformations, which is a well-known effect caused by the hydrodynamic instability associated with elongated contractile particles \cite{marchetti,sriram}. In this regime the contractile stress overcomes the elastic deformations mediated by liquid crystal elasticity and surface tension, eventually leading the vector ${\bf p}$ to fan outwards (or inwards) and the formation of a double vortex triggering the motion (see Fig.\ref{fig1}c,f; Fig.\ref{fig2}b,f; Movie M1). 
Note, in addition, that  the contractile stress induces an effective perpendicular anchoring of ${\bf p}$ at the fluid interface, in agreement with previous works \cite{blow}.
As $|\zeta|$ increases further, the droplet becomes spherical and immotile,  with a topological defect of charge $+1$ at its center (Fig.\ref{fig1}d,g, where ${\cal A}\simeq 2$ for $|\zeta|=6\times 10^{-3}$, and Movie M2). A $-1$ defect would have resulted if the splay deformation of the previous state had pointed inwards. In this configuration, the polar field  exhibits a hedgehog structure with vectors pointing outward while the corresponding flow field displays a symmetric eight-vortex structure, where the fluid is pulled along the vertical/horizontal directions and pushed along the oblique ones (Fig.\ref{fig2}c,g). We note that the existence of this state has also been discussed in Refs.\cite{tjhung1,tjhung3}, where the dynamics of 3d active gel droplet were studied in an unconfined geometry. Interestingly, increasing $|\zeta|$ even further yields a previously unreported motile state, where the droplet acquires a 3d peanut-like shape hosting an integer defect in one of the two bulges (see Fig.\ref{fig1}e,h, where ${\cal A}\simeq 2.5$, and Movie M3). 
This dynamics essentially results from the displacement of the topological defect from the center of the non-motile droplet of Fig.\ref{fig1}c (where flow and polar field are globally symmetric). Indeed, when the defect shifts, the splay pattern is no longer balanced and, for high enough values of $|\zeta|$, the active flow triggers self-propulsion. The flow outside the droplet is somewhat reminiscent of that obtained for the non-motile contractile droplet, although here the velocity is collected longitudinally and expelled along the transversal direction. In contrast, inside the droplet the flow displays four 
vortices causing two jet-like fluids along opposite directions, where the larger one results from higher splay distortions (Fig.\ref{fig2}d,h).   

Such dynamic behavior is summarized in the phase diagram shown in Fig.\ref{fig1}i, where the steady state speed of the center of mass is plotted against the activity. The results are in line with previous works \cite{tjhung1,tjhung2}, except that here we show evidence of a further motile state for large activity. 
In the next subsection we discuss a much less studied system, i.e that in which an three dimensional contractile droplet navigates within a microchannel.

\subsection{Contractile droplet under confinement}

We consider a setup akin to the periodic case, where a droplet of diameter $D=30$ contains a polar liquid crystal and is surrounded by a passive Newtonian fluid. The droplet is placed in the middle of a microfluidic channel, where two parallel flat walls are at distance $L_z$ and periodic boundaries are set along $x$ and $y$. We study two cases that are representative of
a general behavior of the droplet under confinement. More specifically, we investigate the dynamics, by varying $\zeta$, for $L_z$ equal to $105$ and $30$, which give a confinement ratio $\lambda\simeq 3.5, 1$, roughly indicating mild and high confinement. 

\subsection{Mild confinement} If $\lambda\simeq 3.5$, we find that, for low and moderate values of $|\zeta|$ (approximately $|\zeta|\lesssim 7\times 10^{-3}$ where ${\cal A}\simeq 2$), the dynamics is similar to the unconfined case, where the droplet either shows a rectilinear motion or attains a static hedgehog-like configuration. Interestingly, increasing $|\zeta|$ significantly changes the picture.

\begin{figure*}[htbp]
\includegraphics[width=1.0\textwidth]{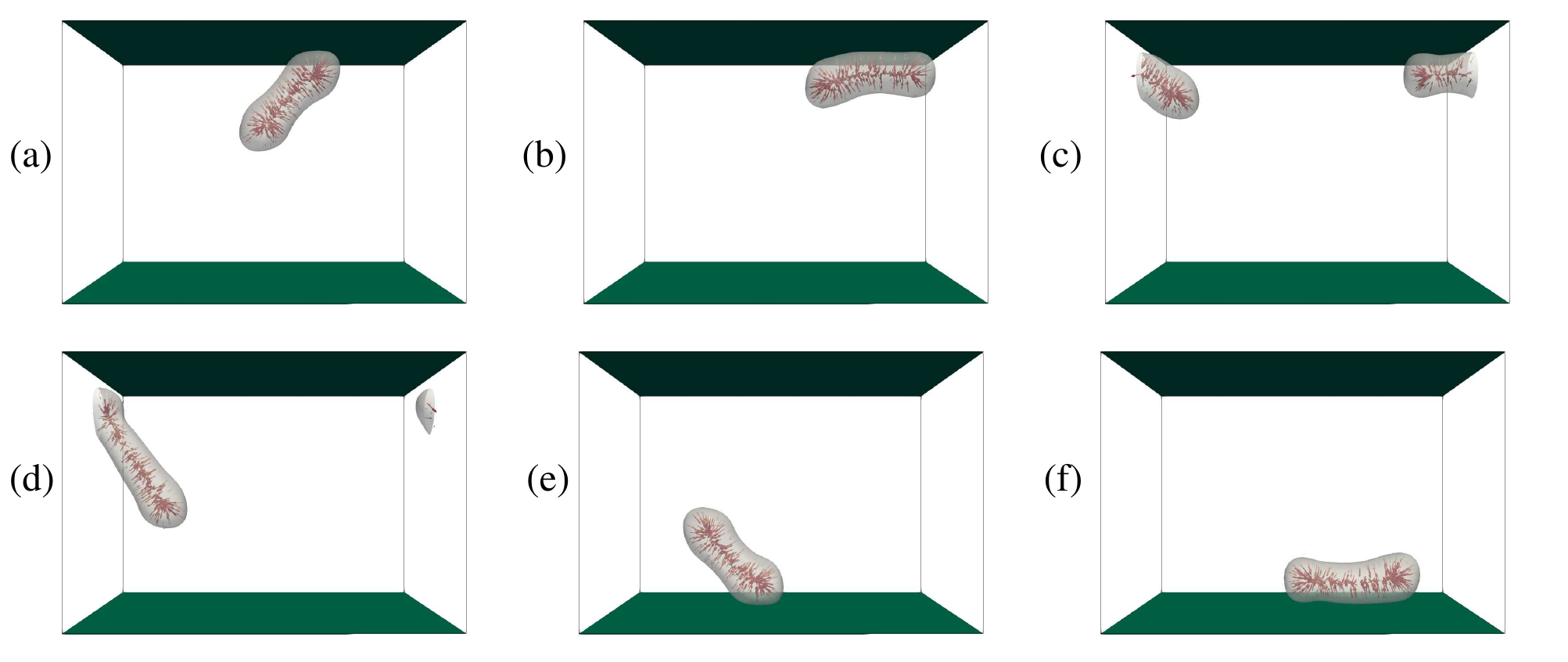}
\caption{Time evolution of an active droplet for $|\zeta|=9\times 10^{-3}$. Snapshots (a), (b), (c), (d), (e) and (f) show the fields $\phi$ and ${\bf p}$.  The droplet acquires motion (for example upwards) following a dynamics akin to that of the peanut-like droplet in Fig.\ref{fig1}e. Afterwards, it bumps against the wall (a) and glides over it (b,c) before turning downwards (d) and colliding/gliding against the opposite wall (e,f). Finally, the process repeats leading to a periodic motion persistent over time. Note that here the polarization points inwards, thus yielding a $-1$ defect at the rear of the droplet.}
\label{fig3}
\end{figure*}

\begin{figure*}[htbp]
\includegraphics[width=1.0\textwidth]{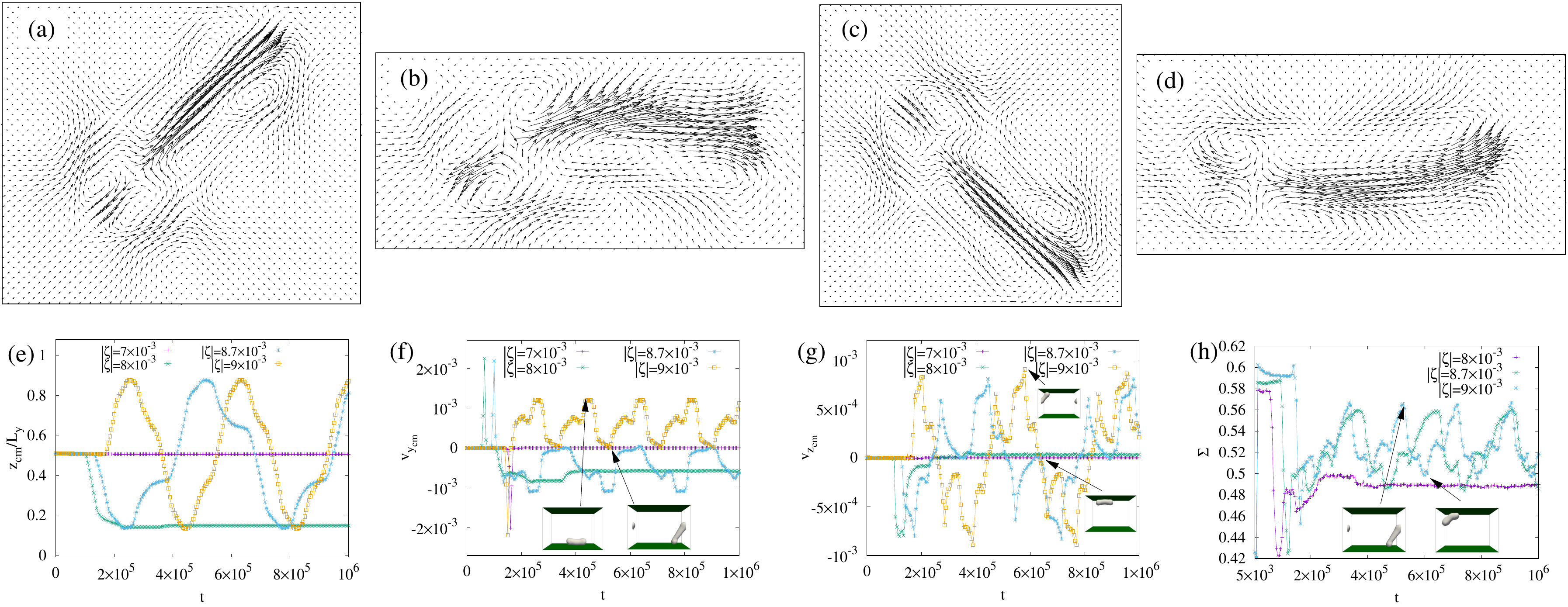}
\caption{(a-d) Section of the fluid velocity along the $y-z$ plane (taken at $x=L_x/2$) for a droplet with $|\zeta|=9\times 10^{-3}$. Far from the walls (a,c), the flow consists of a bidirectional stream, resulting from the splay deformations of the polar field, along the backbone of the droplet. The integer topological defect located at the rear (with respect to the droplet motion) marks the change of direction of the stream. This one is accompanied by four stretched vortices, two large ones at the front and two small ones at the rear. Near the walls (b,d), the momentum sink considerably abates the vortices in their vicinity. This allows the droplet to temporarily glide over the wall and then turn towards the center of the channel. (e) Time evolution of the vertical component ($z$) of the position of the center of mass for four values of activity, $|\zeta|=7\times 10^{-3}$ (no motion), $|\zeta|=8\times 10^{-3}$ (persistent glide over the wall), $|\zeta|=8.7\times 10^{3}$, $|\zeta|=9\times 10^{-3}$ (periodic motion). (f-g) Time evolution of $y$ and $z$ components of the speed of center of mass. (h) Time evolution of the parameter $\Sigma=N_s/N_v$, where $N_s$ is the number of lattice points located at the droplet surface and $N_d$ is the number of lattice points located within, thus proportional to the volume.}
\label{fig4}
\end{figure*}

In Fig.\ref{fig3} we show the time evolution of the droplet and the corresponding velocity field for $|\zeta|=9\times 10^{-3}$ (where ${\cal A}\simeq 3$ , see also Movie M4). The mechanism leading to motility is basically the same as the one discussed for the periodic case, where a peanut-shaped droplet acquires motion because of the unbalance of splay deformations along a preferential orientation. However, in the presence of boundaries, the droplet motion is characterized by periodic hits on opposite walls while moving forward. More specifically, the droplet initially approaches the wall and then glides over it (Fig.\ref{fig3}a,b,c), turns back towards the middle of the channel and finally glides over the opposite wall (Fig.\ref{fig3}d,e,f). 
Although such dynamics looks akin to that discussed in Ref.\cite{tiribocchi_pof2023} for a two dimensional droplet, here the motion is obtained for considerably higher values of $|\zeta|$, and it results from a symmetric four-roll structure of the internal flow field combined with a jet-like flow dragging an integer topological defect (which is absent in 2d). As the droplet approaches the boundaries, the stream of flow along the backbone of the droplet becomes dominant, whereas the vortices near the walls undergo a significant weakening because of the momentum sink. The imbalance between the vortices on both sides of the droplet causes the turn towards the middle of the channel, where a symmetric flow field is restored (see Fig.\ref{fig4}a,b,c,d). The periodic dynamics is also favored by the no-wetting condition, which essentially prevents the droplet from getting stuck at the wall.
We note that an intermediate state between the non-motile and the periodic ones is that found for $|\zeta|\simeq 8\times 10^{-3}$, where the peanut-like droplet swims persistently close to the wall, but at a distance (comparable with the droplet radius $R$) such that the momentum sink effect is reduced. This basically preserves the symmetric structure of the internal flow field and prevents the turn.  

These features can also be captured by tracking position and speed of the center of mass (see Fig.\ref{fig4}e,f,g). The time evolution of the velocity  shows that the maxima of $v_y$ and minima of $v_z$ are observed for a droplet gliding over the wall and approaching the wall, whereas minima of $v_y$ and maxima of $v_z$ are found for a droplet either leaving or approaching the wall. Also, the spikes of $v_y$ at early times capture the initial translation, after which the spherical droplet turns into the peanut-like shape and acquires motion. 

We further note that the oscillations of the trajectory come with periodic shape changes of the droplet during motion.
Such modifications can be tracked by computing $\Sigma=N_s/N_v$, where $N_s$ is the number of lattice points for which 
$0.2\lesssim \phi\lesssim 1.8$, thus accounting for the droplet interface, while $N_v$ is the number of lattice sites where $\phi>1$, which approximates the volume of the droplet.  Since the volume is preserved, this quantity provides a dimensionless number to track surface changes, hence shape deformations, over time. For a spherical droplet, $\Sigma$ can be calculated analytically, and is given by $\Sigma_{sd}=4\pi R^2\xi/\frac{4}{3}\pi R^3=3\xi/R$, where the numerator gives the volume of a shell of thickness $\xi$. If $R=15$ and $\xi\simeq 2$, one has $\Sigma_{sd}\simeq 0.4$. In Fig.\ref{fig4}h we show the time evolution of $\Sigma$ for three values of $|\zeta|$, in cases where either a rectilinear or a periodic motion is observed. Note that while in the former $\Sigma$ is roughly constant at late times
(i.e. shape changes become negligible), in the latter $\Sigma$ oscillates regularly between a state where the droplet approaches the wall (minimal deformation) and another one where the droplet moves towards the middle of the channel (maximal deformation). 
This is because, once the direction of motion has changed, the symmetric flow field has to be restored, an effect that yields the formation of vortices temporarily slowing down and stretching the droplet. Finally, further increasing $\zeta$ leads to droplet rupture. 
 
\subsection{High confinement} Lastly, we mention a few considerations about a regime where $\lambda\simeq 1$, when the size of the droplet is comparable to that of the microchannel.  Note that being the interface width $\xi\simeq 2$ lattice sites, one has $2\xi/L_z\simeq 0.15$, in line with the 2d simulations discussed in Ref.\cite{tiribocchi_sm2023}. 
Under such a high confinement, the immotile regime broadens to $|\zeta|\simeq 6\times 10^{-3}$, essentially because the walls hinder 
a sufficiently high momentum transfer to the surrounding fluid.
For $6\times 10^{-3}< |\zeta|\lesssim 7.5\times 10^{-3}$ the motion is, once again, rectilinear at late times although, unlike the mild confinement regime, it comes after a temporary state where an S-shaped droplet spins around its center of mass (see movie M5 for $|\zeta|=7\times 10^{-3}$).  Its formation results from two splay deformations producing two counter-rotating vortices (thus a torque), and is in agreement with the findings of Refs.\cite{fialho,tiribocchi_sm2023} where it is shown that a combination  of local tangential anchoring of polar field at the poles and normal anchoring at the remaining part of the interface  leads to a similar dynamics for 2d contractile droplets. However, since the perpendicular orientation prevails, the S-shaped droplet lasts for a short period of time and the two branches merge into a single one hosting a single splay deformation.  As in the case  of mild confinement, increasing $\zeta$ leads to droplet breakup. 

\section{Conclusions}
In this work, we have numerically studied the dynamics of a  three-dimensional contractile droplet in systems with periodic boundaries and within microchannels of different sizes. 
In the former, our results confirm the shapes and 
dynamic regimes discussed in previous works \cite{tjhung1,tjhung3}, and unveil the existence of a further motile state at high activity characterized by a peanut-shaped droplet (hosting an integer topological defect) propelled by a combination of a jet-like flow and four fluid vortices located in the droplet interior. 
Since this state precedes the droplet breakup, it would be of interest to investigate to what extent it may provide a minimal model for cytokinesis, where an actomyosin ring generates contractile forces, constricts the cell and drives its division \cite{synth_bio,lopez_natcomm}. 
Within a microchannel, the confinement ratio significantly affects the motion. Under mild confinement, the droplet exhibits either a rectilinear motion for low/moderate values of $|\zeta|$ or a unidirectional oscillatory one, with repeated hits on opposite walls, for larger values of $|\zeta|$, where it acquires a peanut-like shape. A similar dynamics has been experimentally observed in Ref.\cite{Maass_natcomm}, where an active droplet navigates upstream following an oscillating trajectory.
Under high confinement, the momentum sink strongly inhibits the momentum transfer, especially for low and intermediate values of activity where the droplet is motionless. At higher values, a rectilinear motion is generally found at late times. 
Here, oscillations of the trajectory are absent basically because of the lack of sufficient space, although in the presence of larger shape deformations (achieved by further increasing $|\zeta|$) combined with a unidirectional motion, they could be observed in principle. We note that, besides the mechanism leading to motility discussed in this work, under high confinement spontaneous motion could be triggered by alternative sources, such as active self-advection \cite{tjhung1,tjhung2,liverpool_soft,aranson1,aranson3},  tractionless tank-treading and, potentially, capillary force\cite{liverpool_prl,liverpool_soft}. It would be of interest to study the effect of these sources of driving in three dimensional devices as well as with other boundary conditions, such as partial slip or different wetting, which are expected to considerably impact the dynamics and the present picture. Of particular relevance for applications is the study of the physics of many self-organized and interacting active droplets \cite{aranson2}, where purposefully designed high performance simulations incorporating hydrodynamics are often mandatory \cite{tiribocchi_light,tiribocchi_prep}.
A final remark concerns the limitations of the approach adopted in this work in providing a more accurate description of cells or their biological analogs. Besides the approximation of single viscosity (which could be in principle relaxed by assuming $\eta$ dependent on $\phi$), the lack of nucleus and distribution of the contractile stress in the whole droplet, rather than in a cortex surrounding the nucleus,  make the active droplet a simplified model. From a purely mechanical standpoint, an active double emulsion could provide a more realistic model, where the nucleus would be represented by the inner droplet and the actomyosin would be confined in the layer between inner and outer droplets  \cite{negro_natcomm}. Also, our model does not capture interface (or membrane) fluctuations, which could be included, for instance, through a noise term or by advanced computational fluid dynamics techniques based on grid refinement methods.

\acknowledgments
A.T. and M.L. acknowledge  the support of the GNFM-INdAM. A.T., M.L. and S.S. acknowledge the support of ERCPoC Grant No. 101187935 (LBFAST).

\bibliography{biblio} 
\bibliographystyle{unsrt}

\end{document}